\documentclass[journal]{IEEEtran}
\ifCLASSINFOpdf
  \usepackage[pdftex]{graphicx}
\else
\fi
\usepackage[cmex10]{amsmath}

\usepackage{mdwmath}
\usepackage{mdwtab}

\usepackage{amsfonts}
\usepackage{algorithm}
\usepackage{algpseudocode}
\usepackage{lipsum}

\usepackage{caption}

\begin{document}
%
\title{GPU Accelerated Fractal Image Compression for Medical Imaging in Parallel Computing Platform}
%
%
%

\author{Md.~Enamul~Haque, \IEEEmembership{}
        Abdullah~Al~Kaisan, \IEEEmembership{}
        Mahmudur~R~Saniat, \IEEEmembership{}
         and~Aminur~Rahman \IEEEmembership{}

\thanks{Md. Enamul Haque is with King Fahd University of Petroleum \& Minerals, Dhahran, 31261, Kingdom of Saudi Arabia. (phone: +966-50-2389368, email: g201204920@kfupm.edu.sa, enamul\_cse@yahoo.com)}
\thanks{Abdullah Al Kaisan is with King Fahd University of Petroleum \& Minerals, Dhahran, 31261, Kingdom of Saudi Arabia. }
\thanks{Mahmudur R Saniat is with Tilburg University,Warandelaan 2, 5037 AB Tilburg, Netherlands}
\thanks{Aminur Rahman is with  Banglalink Digital Communications Limited, Dhaka, Bangladesh}}

\maketitle

\begin{abstract}
In this paper, we implemented both sequential and parallel version of fractal image compression algorithms using CUDA (Compute Unified Device Architecture) programming model for parallelizing the program in Graphics Processing Unit for medical images, as they are highly similar within the image itself. There are several improvement in the implementation of the algorithm as well.
Fractal image compression is based on the self similarity of an image, meaning an image having similarity in majority of the regions. We take this opportunity to implement the compression algorithm and monitor the effect of it using both parallel and sequential implementation. Fractal compression has the property of high compression rate and the dimensionless scheme. Compression scheme for fractal image is of two kind, one is encoding and another is decoding. Encoding is very much computational expensive. On the other hand decoding is less computational. The application of fractal compression to medical images would allow obtaining much higher compression ratios. While the fractal magnification an inseparable feature of the fractal compression would be very useful in presenting the reconstructed image in a highly readable form. However, like all irreversible methods, the fractal compression is connected with the problem of information loss, which is especially troublesome in the medical imaging. A very time consuming encoding process, which can last even several hours, is another bothersome drawback of the fractal compression.

\end{abstract}


\IEEEpeerreviewmaketitle


\section{Introduction}

\IEEEPARstart{T}{HE} advent of modern GPUs for general purpose computing has provided a platform to write applications that can run on hundreds of small cores. CUDA (Compute Unified Device Architecture) is NVIDIAs implementation of this parallel architecture on their GPUs and provides APIs for programmers to develop parallel applications using the C programming language. CUDA provides a software abstraction [1] for the hardware called blocks which are a group of threads that can share memory. These blocks are then assigned to the many scalar processors that are available with the hardware. Eight scalar processors make up a multiprocessor and different models of GPUs contain different multiprocessor counts.

Compression is an important and useful technique that can be slow on current CPUs. We wanted to investigate ways to parallelize compression specifically on GPUs where hundreds of cores are available for computation. By investigating three forms of compression (JPEG, ogg, zip), we can characterize what components and forms of compression are suited to parallelization. Computers store up images as collections of bits representing pixels or points forming the picture elements. Numerous pixels are essential to store moderate quality images, because the human eye can process bulky amounts of information (some $8$ million bits). Most data contain a little amount of redundancy, which can sometimes be aloof for storage and replaced for recovery, but this redundancy does not lead to high compression ratios. The typical methods of image compression come in more than a few varieties. The existing most accepted technique depends on removing high frequency components of the signal by storing only the low frequency components (Discrete Cosine Transform Algorithm). JPEG (still images), MPEG (motion video images), and H$.261$ (Video Telephony on ISDN lines) compression algorithms mainly use this scheme.

As,we know huge quantity of medical images are produced everyday and those are to be stored in some place for future reference and usage. It takes a lot storage and time to store them. Thus we show that if we use GPUs to process the image and reduce the size for storing, it makes sense. There may some questions arise that why to consider medical images only. The main reason behind this is medical images are self similar and can be compressed easily with better size reduction [18]. It is described in more detail in the later sections.

On the other hand, images except medical areas, most of them are not black and white. They are complex in nature meaning the image contain more information compared to black and white. Thus it is not very beneficial to compress this sort of images. We will not get much reduction of the size after compression as we need to put most of the information intact to get back the original image [19].

Fractal image compression [2] method will encode an image as a set of transforms that are very analogous to the copy machine scheme. For example, the branch has detail at each level, the figure reconstructed from the transforms too. The image can be decoded at any size since it has no usual range. The additional element required for decoding at larger scale is generated involuntarily by the encoding transforms. We could decode an image of an individual escalating the size with each iteration, and ultimately can observe skin cells or perhaps atoms; it may be surprising if this detail is real. The response is in fact no. At what time the image was digitized, the detail is not at all associated to the actual detail present; it is just the artifact of the encoding transforms which initially only encoded the large- scale features. Nonetheless, in some occasion the detail is pragmatic at low magnifications, and this can be handy in Security and Medical imaging applications.

As we know that the compression is based on finding the similar fragment and encoding them accordingly, thus there is redundant processing involved. So, the processing time becomes less if these redundant processing can be deployed to number of processors. This can be done with the help of parallel graphics processors. Hence, CUDA comes in the way to perform parallel processing. 

The remainder of this paper is organized as follows. In section II, we mentioned a concise overview of the related work done in this arena. We explained fractal image compression briefly in section III. In section IV, we describe our experimental setup mentioning how we prepared our images, CPU and GPU configuration. In section V, we demonstrate our implementation details. In section VI, we demonstrate our results and performance evaluation. Finally in section VII, we conclude and suggest future direction of our work.


\section{Related Work}

In this section, several contemporary research related to fractal image compression in parallel computing platforms will be explored. Authors have presented different aspect of their implementation and the performance comparison. There are several works related to fractal image compression from previous authors. They proposed different approaches for the solution. We will be having some of their ideas to get the difference from our approach.

Wakatani et al. [3] explained about implementing adaptive fractal image coding algorithms on GPUs by using CUDA in order to achieve a given quality of a compressed image and evaluate the improved version of the coding algorithm by enhancing the occupancy of GPU cores. The adaptive fractal image coding algorithm consists of three two- fold loops which include no data dependences, so it is very easy to parallelize the coding algorithm by distributing the loop elements on multiple computing cores. Original implementation of the adaptive fractal image coding on GPU is that the speedup using the range block size of $16\times16$ is worse than that of $4\times4$ and $8\times8$ range blocks due to the hardware requirement like the size of the shared memory. Improved implementation: The occupancy should be enhanced by adding more threads for each thread block.

Franco mentioned about [4] improving Taking advantage of the NVIDIA Tesla C$870$ that compiles with the CUDA model to improve the execution time of the 2D fast wavelet transform. The author implemented in this way, 1D FWT is applied to a row of an image of $n$X$n$ pixels, $\frac{n}{2}$ low pixels and $\frac{n}{2}$ high pixels are created.In the last Stage, $\frac{n}{2}$-th low pixel and $\frac{n}{2}$-th high pixel depended on the $(n-1)$th, $n$th, first and second original pixel. A .pgm file is read into a page-locked main memory buffer. The main memory buffer is copied to a global memory buffer. The 1D-FWT is applied to each row of the image. The image matrix is transposed again to recover the initial data layout. The resulting image is copied back from the global memory buffer to the page-locked main memory buffer.

Simek [5] mentioned that diagnostic materials are increasingly acquired in a digital format. The common need to daily manipulate huge amount of data brought about the issue of compression within a very less stipulated amount of time. Attention is given to the acceleration processing flow which exploits the massive parallel computational power offered by NVIDIA GPU.

Initially, CUDA was used to perform wavelet decomposition of an image,floating point matrix multiplications, additions and calls to CUBLAS library. Then he imported that code to CUDA. In step-II, the wrote MEX file to call the code to perform the decomposition and thresholding the wavelet coefficient. Matlab code was running in a double precision mode. Data is transferred into single precision mode before transferring into CUDA. Computation on the GPU is single precision arithmetic and the result is transformed back to double precision before it is returned into MATLAB.

Lin [6] presents a CUDA-based implementation of an image authentication algorithm with NVIDIAs Tesla C$1060$ GPU devices. They showed the comparison with the original implementation on CPU, our CUDA-based implementation works $20$x-$50$x faster with single GPU device. And experiment shows that, by using two GPUs, the performance gains can be further improved around $1.2$ times in contrast to single GPU.

Park et al. [7] discussed about the vital issues in plan and evaluation of image processing algorithms on the colossal parallel graphics processing units (GPUs) using the compute unified device architecture (CUDA) programming model. Customized for image processing, a set of metrics is anticipated to quantitatively estimate algorithm characteristics. Moreover, it is illustrated a series of image processing algorithms map readily to CUDA using multi view stereo matching, linear feature extraction, JPEG$2000$ image encoding, and non photorealistic rendering (NPR) as our paradigm applications. We circumspectly select the algorithms from major domains of image processing, so they intrinsically include a diversity of sub-algorithms with varied characteristics when implemented on the GPU. On the basis of execution time, performance is evaluated and is compared to the fastest host-only version implemented using OpenMP. Furthermore, the observed speedup varies broadly depending on the characteristics of every algorithm. 

Erra et al. [16] used GeForce FX $6800$ graphics card for the fractal image processing and found significant improvement over CPUs. They used image with resolution $256 \times 256$ pixels and range block size of $4 \times 4$. The domain has to be twice the size of range, and in their case it was $15,625$ elements in total. Brute force strategy performs total $4,096 \times 15,625 = 64,000,000$ possible testing operations. For this CPU took $280$ seconds and GPU took $1$ second only. 

Sodora et al. [17] implemented both the encoder and decoder in C++ and ran on intel Core 2 Duo P$7450$ processor with each cores clock speed of $2.13$ GHz. They implemented the parallel version of the program in C++ and OpenCL library and ran on Nvidia $9600$M GS. The results showed that fractal compression is useful as a resolution-independent video capture format.

\section{Fractal Image Compression}
Fractal is defined with fractal geometry [13] which refers to geometric shapes that can be subdivided into smaller parts similar to the original fragment. Mandelbrot [13] mentioned the function $f(z_n) = z_n^2 + z_0$ which explains the idea of the similar fragment of the original image. The basic idea behind fractal image compression is to reduce the redundant pixels, thus reducing the storage space, from the original image without loosing the structure. There are several examples of fractal geometry that are, in some way, similar to the whole. For example, Cantor set and von Koch curve are made up of similar copies. The middle third Cantor set [14] is made up of the union of two similar copies of itself leaving the middle third part out. This process is done infinitely, where the $n$-th set is described as

\begin{equation}
C_n=\frac{C_{n-1}}{3} \cup \bigg( \frac{2}{3} + \frac{C_{n-1}}{3} \bigg)
\end{equation}

\subsection{Iterated Function System}
Iterated function system is one of the key point for finding the self similarity of from the segments. We need to know about the contraction before going in detail about the iterated function system. If $D$ is a closed subset of $\mathbb{R}^n$, where it is both of them are almost similar. \textit{Contraction} is defined with some mapping between $S$ and $D$, $S:D \rightarrow D$. This mapping is called contraction on $D$ if there is a number $c$ with $0<c<1$ so that $|S(x)-S(y)| \leq c|x-y|$ for all $x,y \in D$. Every contraction is continuous. When $|S(x)-S(y)| = c|x-y|$, then $S$ becomes geometrically similar, and called \text{contracting similarity}. When $m \geq 2$ for a finite set of contractions ${S_1, S_2, ... S_m}$, then this set is called an iterated function system. Also, any non-empty compact subset $F$ of $D$ is called an \textit{attractor} or \textit{invariant set} of the IFS when 

\begin{equation}
F = \bigcup_{i=1}^{m}S_{i}(F).
\end{equation} 

IFS determines one specific property which is to find a unique attractor.For simplicity, if $F$ is taken as the middle third Cantor set and $S_1, S_2: \mathbb{R} \rightarrow \mathbb{R}$, then 

\begin{equation}
S_1(x) = \frac{1}{3}x
\end{equation} 

\begin{equation}
S_2(x) = \frac{1}{3}x+\frac{2}{3}
\end{equation} 

We can determine from the above equations that $S_1(F)$ and $S_2(F)$ are the right halves of $F$ where $F=S_1(F) \cup S_2(F)$. So, The IFS has an attractor $F$ with contractions ${S_1,S_2}$

\subsection{Affine Transformation}
Affine transformation of an image refers to random combination of rotation, scaling, skewing or translation. The matrix equation for affine transformation is given below,

\begin{equation}
x^{'} = Ax+d
\end{equation} 

Where $x$ is co-ordinate vector, $A$ is image matrix and $d$ is displacement vector.

\subsection{Partitioned Iterated Function System}

Partitioned iterated function system restricts transformation on specific subsets of input domain blocks. We get range blocks after transforming these domain blocks. We partition the image into several range blocks and find domain and transformation for each of these range blocks. The transformation maps the domain blocks onto the range blocks. We have found how to transform the points of the original image into new sets of points, but this is not the only transformation while compressing. We need to consider the image intensity as well to define the original image along with the position. We considered the brightness of a pixel as a function of its position. For example, if the position of a pixel is $(x,y)$, then the brightness, $z=p(x,y)$.

Another step is to scale and offset the brightness value as part of the transformation. Hence, the PIFS transformation domain becomes, 

\begin{equation}
w_i \begin{bmatrix}
       x         \\[0.3em]
       y \\[0.3em]
       z
     \end{bmatrix}
    = \begin{bmatrix}
       a_i & b_i & 0        \\[0.3em]
       c_i & d_i & 0 	 \\[0.3em]
       0 & 0 & s_i
     \end{bmatrix}
     \begin{bmatrix}
       x         \\[0.3em]
       y \\[0.3em]
       z
     \end{bmatrix}
     +
     \begin{bmatrix}
       e_i         \\[0.3em]
       f_i 	   \\[0.3em]
       o_i
     \end{bmatrix}
\end{equation}

where, $s_i, o_i, w_i$ are scale, offset, and transformation respectively.

\subsection {Collage Theorem}

Now, we need an efficient way to find the $w_i$. Collage theorem [15] helps to find this $w_i$ effectively. Collage theorem states that,

\begin{equation}
h(f,W(f)) \leq \epsilon \Rightarrow h(f,A) \leq \frac{\epsilon}{1-\epsilon}
\end{equation}

where, $A$ is the attractor of the IFS, $h$ is the Hausdorff metric. This metric finds the optimum attractor which is more closer from the available set of transformations. This transformation also limits the infinite searching among the combinations of the attractors.

\section {Experimental Setup}
We have used the below configured machines for our experiment. Table-I shows the CPU configuration which was used while processing the image without any parallelization. Table-II shows the GPU configuration where our CUDA code was run.

The below components are used for the experiment:

\begin{enumerate}
\item Three sizes of input image $(256 \times 256, 512 \times 512$, and $1024 \times 1024)$.
\item Three kernel block layouts $8\times8$, $16\times16$, and $32\times32$.
\item Three decompression output image sizes, (smaller than, equal to, and grater than original compressed image size)
\end{enumerate}

We have chosen fractal images for both compression and decompression implementation. Initially the image is copied from file to GPU memory (Main memory and Texture Memory). This is done using FreeImage library. This is a open source image library written in C++ (C++ wrapper). This library has been designed to be extremely simple in use. It has unique FreeImageIO structure which makes it possible to load images from virtually anywhere. Possibilities include standalone files, memory, cabinet files and the internet, all this without recompiling the library.

Once the image is loaded to the GPU memory, our encoding kernel is invoked. Each thread is assigned to one range block. After computation of the encoding, the result is saved back to the encoding file. Table-I shows the CPU configuration used for our experiment.


\begin{table}[ht]
 \caption{CPU Configuration} 
 \label{table_example}
 \centering 
 \begin{tabular}{c c} 
 \hline\hline 
 \textbf{Property} & \textbf{Value} \\ [0.5ex] 
 \hline 
CPU Clock rate                            & $2.30$ GHz\\
L1 D-Cache 				& $64$ KB \\
L1 I-Cache 				& $64$ KB \\ 
L2 Cache 					& $512$ KB \\
Memory Size 				& $2$ GB \\ [1ex] 
 \hline 
 \end{tabular}
 \end{table}

Table-II shows the GPU configuration used for the experiment. 

\begin{table}[ht]
 \caption{GPU Configuration} 
 \label{table_example}
 \centering 
 \begin{tabular}{p{4 cm}  p{4 cm}} 
 \hline\hline 
 \textbf{Property} & \textbf{Value} \\ [0.5ex] 
 \hline 
CUDA Driver Version / Runtime Version  &  $5.0 / 5.0$ \\ 
CUDA Capability Major/Minor version number & $2.0$ \\
Total amount of global memory & $4096$ MBytes \\
(16) Multiprocessors x ( 32) CUDA Cores/MP: &    $512$ CUDA Cores\\
GPU Clock rate:                              &   $1301$ MHz ($1.30$ GHz)\\
  Memory Clock rate:                           &   $1701$ Mhz\\
  Memory Bus Width:                            &   $384$-bit\\
  L2 Cache Size:                               &   $786432$ bytes\\
  Max Texture Dimension Size (x,y,z)           &   $1$D=($65536$), 2D=($65536,65535$), 3D=($2048,2048,2048$)\\
  Max Layered Texture Size (dim) x layers       &  $1$D=($16384$) $\times 2048$, $2$D=($16384,16384$) $\times 2048$\\
  Total amount of constant memory:              &  $65536$ bytes\\
  Total amount of shared memory per block:      &  $49152$ bytes ($48$ KBytes)\\
  Total number of registers available per block: &  $32768$\\
  Warp size:                                     & $32$\\
  Maximum number of threads per multiprocessor:  & $1536$\\
  Maximum number of threads per block:          &  $1024$\\
  Maximum sizes of each dimension of a block:   &  $1024 \times 1024 \times 64$\\
  Maximum sizes of each dimension of a grid:    &  $65535 \times 65535 \times 65535$ \\
  Maximum memory pitch:                         &  $2147483647$ bytes\\
  Texture alignment:                            &  $512$ bytes\\
  Concurrent copy and kernel execution:        &   Yes with 2 copy engine(s)\\
  Run time limit on kernels:                    &  No\\
  Integrated GPU sharing Host Memory:          &   No\\
  Support host page-locked memory mapping:     &   Yes\\
  Alignment requirement for Surfaces:           &  Yes\\
  Device has ECC support:                       &  Disabled\\
  Device supports Unified Addressing (UVA):    &   No\\
  Device PCI Bus ID / PCI location ID:          &  $12 / 0$\\
 \hline 
 \end{tabular}
 \end{table}

\section{Implementation}

In this section, we will be discussing about the implementation details of the program. Fractal image abstraction [8] tries to get as much information as possible while encoding. In this way the decompress image becomes more accurate to the original one. We observed that from our approach there are infinite number of possible transformation to be experimented. We wanted to limit this infinite number to some tolerable extent so that the running time does not affect the performance. We have chosen eight transformation functions for this purpose.

The input image is divided into $n \times n$ range blocks, R and overlapping domain blocks, $D$ $2n \times 2n$. Eight affine transformation functions are used with the image.
The set of transformations are symmetrical in our case as we just considered rotations ($90,180,270$ degree) and mirroring. When the affine transformation is applied on any square block the resultant block becomes square as well.
We need to map the domain onto a range block considering that the range size is half the size of a domain. So, we needed to calculate contraction [9] information from the distance between them. The relation between domain and range is fixed, so the contracting transformation is also fixed.

\begin{figure}[!h]
  \centering           
        \fbox{\includegraphics[width=0.21\textwidth]{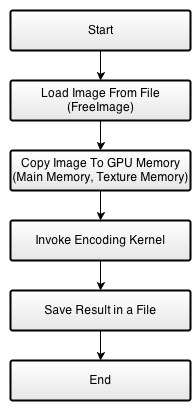} }  
        \caption{General flow of implementation}         
\end{figure}

The domain size is used twice as of the size of range to measure the contracting transformation. Least square method [10] is used for calculating contractivity. The Method of Least Squares is a procedure to determine the best fit line to data. The proof uses simple calculus and linear algebra. The basic problem is to find the best fit straight line $y = ax+b$ given that, for $n_1, ..., N,$ the pairs $(x_n, y_n)$ are observed. The method easily generalizes to finding the best fit of the form
Collage Theorem [11] characterizes an iterated function system whose attractor is close. It is calculated to find the best fit domain and transform function. The major steps for implementing the fractal compression on GPU consists the below ones.

\begin{itemize}
  \item Codebook generation
  \item Least Squares
  \item Symmetry operation
\end{itemize}

The size of the codebook is a determining factor in the computational requirement of the algorithm much previous work has been put into reducing its size, whilst still getting a good compression [3]. However many of these techniques do not lend themselves to parallelization. It is not necessary to pre-calculate the codebook it can be done on the fly. The codebook used in this implementation are generated on the fly. In general not all domains of the desired size are considered, it is common to restrict the search to a grid. The spacing in this grid is referred to as STEP. 

\begin{algorithm}
  \caption{Codebook Generation}\label{codebook}
  \begin{algorithmic}[1]
  \For {$x=0 \to STEP$}
   \For {$y=0 \to STEP$}
     \State $c \gets [x, x+2n] [y, y+2n]$	
      \For {$i=1 \to 8$}
      	\State add $S_i(\tau(c))$
      \EndFor
    \EndFor
   \EndFor
  \end{algorithmic}
\end{algorithm}

We get a metric for a good match after utilizing the least squares method (Algorithm 2) to compare a code block and a range block, and we are able to calculate $s$ and $o$, the scaling and offset of the pixel brightness. The metric will be the residuals, if a code block $a$ is to be transformed into range block $b$, we have to minimize the value of $R$, where $R$ is computed from the below equation.

\begin{equation}
R= \sum \limits_{i=1}^{n} (s.a_i + o.-b_i)^2
\end{equation}

$a_i$ defines the $i$-th pixel in the code block from domain and $b_i$ is the $i$-th pixel from the range we want to map. $s$ and $o$ defines the regression line and are the scaling and offset we will need to apply to transform pixel intensities when decompressing an image. 

\begin{equation}
s= \frac{n \Big( \sum \limits_{i=1}^{n} a_i b_i\Big) - \sum \limits_{i=1}^{n} a_i \sum \limits_{i=1}^{n} b_i }{n \sum \limits_{i=1}^{n} a_i^2 - \sum \limits_{i=1}^{n} a_i}
\end{equation}

\begin{equation}
o= \frac{\sum \limits_{i=1}^{n}b_i - s \sum \limits_{i=1}^{n} a_i}{n}
\end{equation}

If the variance $n \sum \limits_{i=1}^{n} a_i^2 - \big(\sum \limits_{i=1}{n}a_i\big)^2 = 0$ of the domain block, then the slope of the regression line is zero and the offset is a simple average. Then equation (9) and (10) can be replaced with (11) and (12). The variance of the range block can be used to determine if it is a shadow block. If there is no variance, it will only map to domain blocks with no variance. We do not need to search for a domain block for a shadow block, if we do this then we do not need to process code blocks with zero variance.

\begin{equation}
s=0
\end{equation}

\begin{equation}
o = \frac{\sum \limits_{i=1}^{n}b_i}{n}
\end{equation}

\begin{algorithm}
  \caption{Least Square}\label{least square}
  \begin{algorithmic}[1]
	\For {\textbf{all} $d \in D$}
		\For {\textbf{all} $b \in R$}
			\For {$i=1 \to 8$}
				\State $a \gets S_i (\tau(d))$
				\State ${s,o} \gets \textbf{getLeastSquares}(a,b)$
				\State $R=\textbf{computeResiduals}(a,b,s,o)$
				\If {$R < r_j.R$}
					\State $r_j.{s,o} \gets {s,o}$
					\State $r_j.domain \gets d$
					\State $r_j.S \gets i$
				\EndIf
			\EndFor
		\EndFor
	\EndFor
  \end{algorithmic}
\end{algorithm}

The key observation when parallelizing the algorithm is that launching a thread for each range block will avoid write conflicts. This leads to a simple parallel version (Algorithm 3). A lossy and badly compressed image taking $1$ minute on the CPU, would take $0.5$ second with this naive kernel. The kernel utilizes CUDA's texture fetching to load range and domain data. Naively created kernels for the CUDA framework might boost performance compared to a CPU implementation, but to fully realize the potential of the GPU, the code must be tailored to fit the hardware. There are implementations for the two methods simple and quad$4$, optimization efforts has focused on the quad version. In order to focus on the algorithms and not calculating indexes, all input images are assumed to be square and have dimensions $2i$. In general, Algorithm 3 is the parallel version of Algorithm 2 and implemented on CUDA platform.

\begin{algorithm}
  \caption{Parallelization}\label{gpu}
  \begin{algorithmic}[1]
  \State $b \gets r_{thread\_index}$
	\For {\textbf{all} $d \in D$}
			\For {$i=1 \to 8$}
				\State $a \gets S_i (\tau(d))$
				\State ${s,o} \gets \textbf{getLeastSquares}(a,b)$
				\State $R=\textbf{computeResiduals}(a,b,s,o)$
				\If {$R < r_j.R$}
					\State $r_j.{s,o} \gets {s,o}$
					\State $r_j.domain \gets d$
					\State $r_j.S \gets i$
				\EndIf
		\EndFor
	\EndFor
  \end{algorithmic}
\end{algorithm}

\section{Experimental Results and Performance Evaluation}

Initially the image is broken into ranges and domains. Ranges are non overlapping, meaning every ranges are distinct from each other. Domains are overlapping and they are twice the size of a range. For each range location, all the domains are tested with affine transformation to get the most accurate new range block. All the domains and each range transformations are differentiated using least square method to get the difference. The set of domain and transformation which gave the minimum difference is considered as the best possible solution for that range location.As, the combinations can be unlimited, we considered eight transformations (rotations, mirroring, scaling)
The execution time shows best result when the kernel dimension is $16\times16$ for different image size using global memory.

\begin{figure}[!h]
  \centering           
        \fbox{\includegraphics[width=0.47\textwidth]{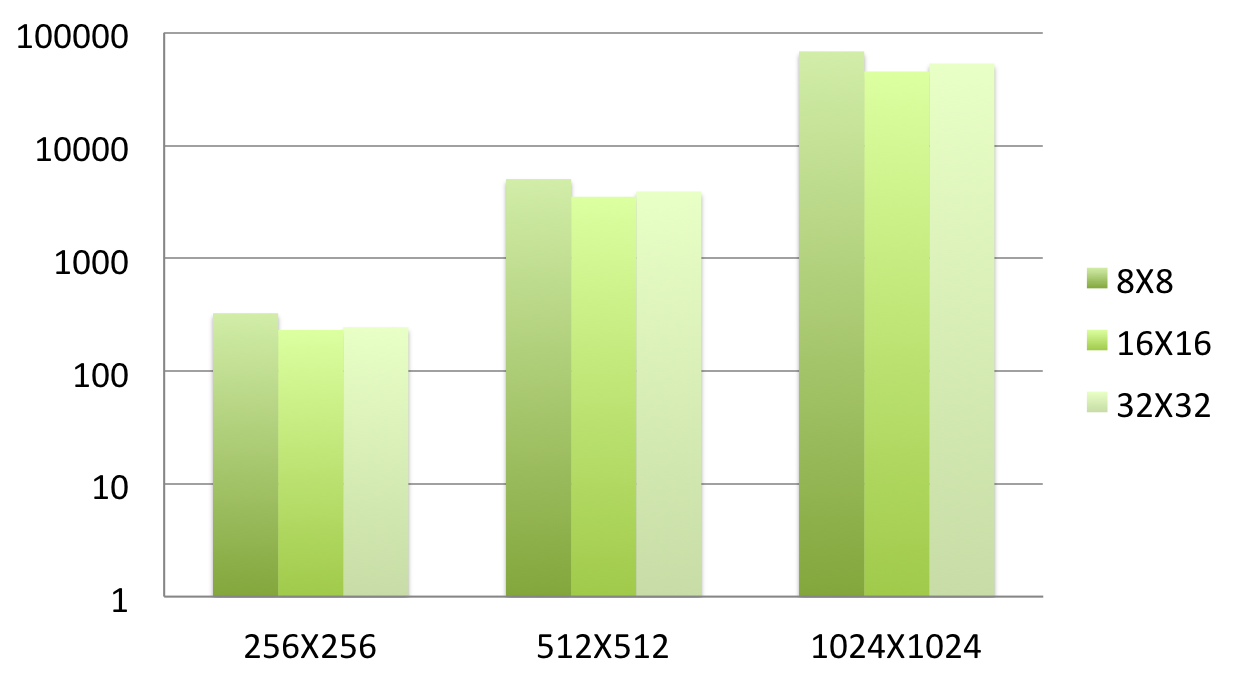}}   
        \caption{ Image Size Vs. Execution Time (ms) (Shared Memory)}         
\end{figure}

The below graph shows the execution time for different image sizes when used in texture memory.

\begin{figure}[!h]
  \centering           
        \fbox{\includegraphics[width=0.47\textwidth]{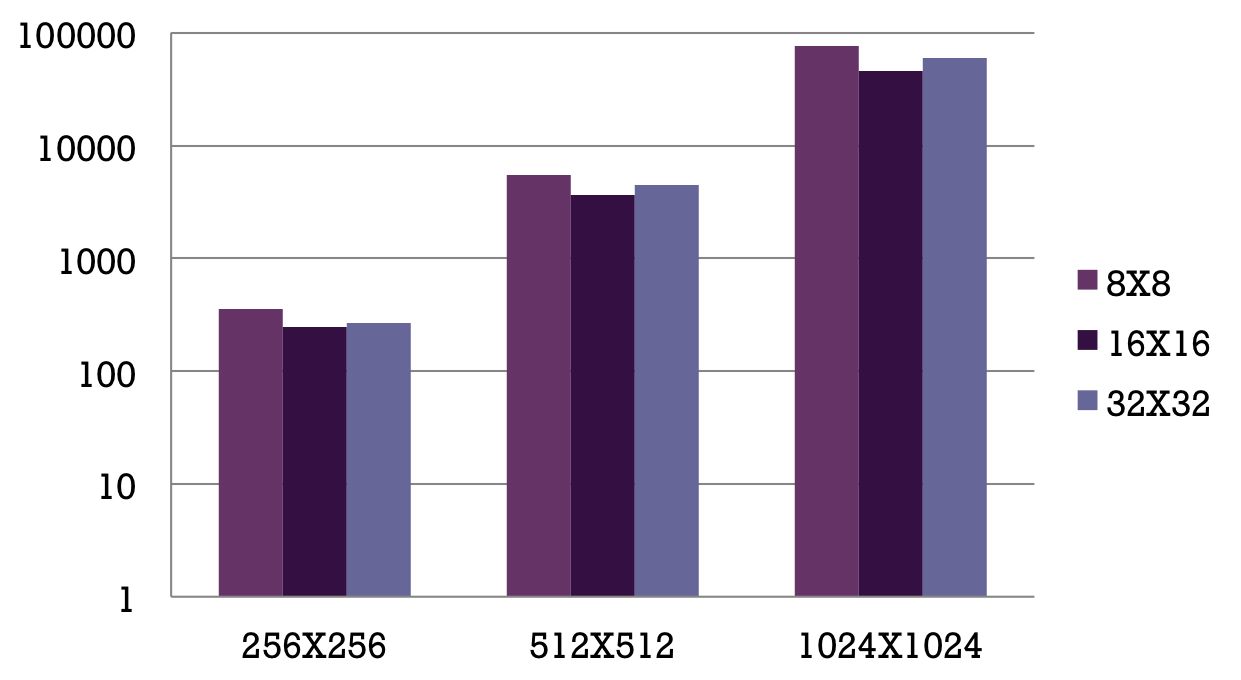}}   
        \caption{Image Size Vs. Execution Time (ms) (Texture Memory Implementation)}         
\end{figure}

From our experimental result (Table-III) it shows that the speedup in execution time affects greatly in GPU compared to CPU. Also, the image size reduction is more than 40\% in all cases. The compression rate depends on the choice of rangeÕs size and the interleaving size of domains and to what extends similarities exists in the image. This result was extracted when using $4\times4$ range size with $4\times4$ interleaved domain. The larger range and less interleaving domains give better compression rate but result in poor decoding quality. The compression rate is not limited to the listed number. This is only when decompressing to the same size, when for example decompressing to $4$ times the original size the 40\% is now 85\%.

\begin{table*}
 \caption{Speedup} 
\begin{center}
\begin{tabular}{lllll}\hline
Image Size  & CPU Execution Time & GPU Execution Time & Speedup & Size reduction\\ 
\hline
$256 \times 256$ & $247.432$ & $0.232$ & $1066.517241$ & $40$ \%\\
$512 \times 512$ & $2369.4$ & $3.521$ & $72.9593865$ & $40$ \% \\
$1024 \times 1024$ & $39276.5$ & $45.407$ & $864.9877772$ & $41$ \% \\
\hline
\end{tabular}
\end{center}
\label{tab:1}
\end{table*}

\begin{figure}[!h]
  \centering           
        \fbox{\includegraphics[width=0.47\textwidth]{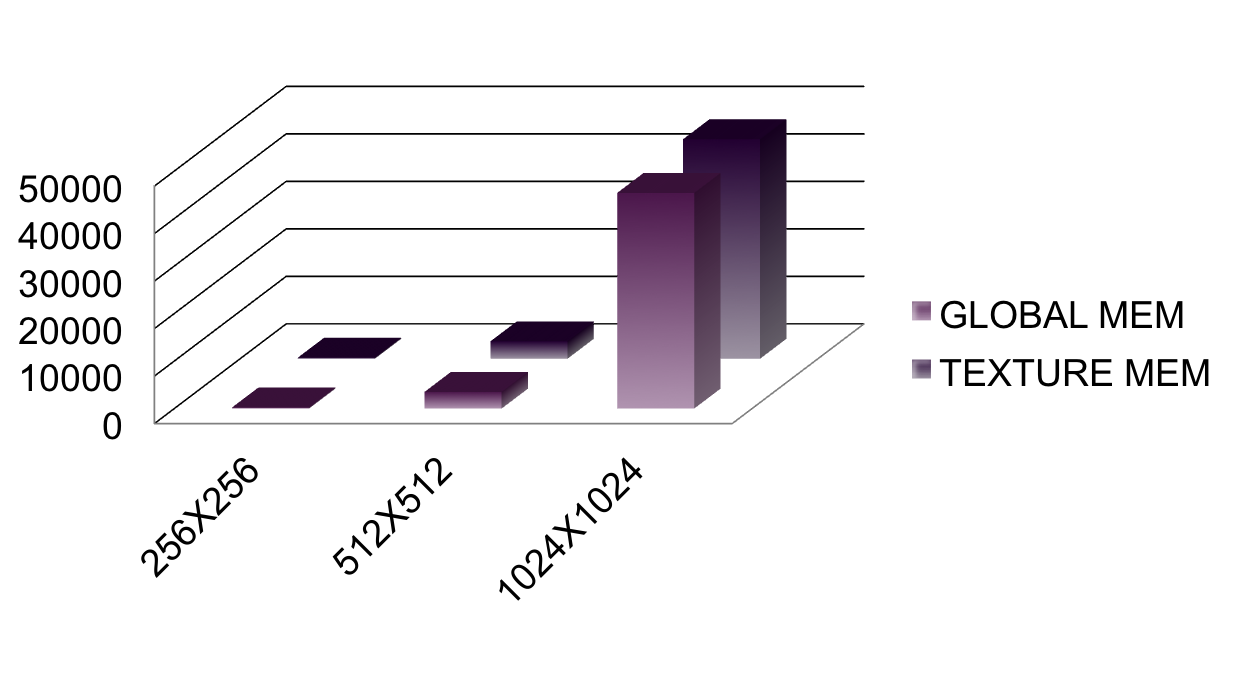}}   
        \caption{Comparison between Global and Texture Memory in execution time}         
\end{figure}

Figure 5 depicts the detail specification of the implementation of the fractal algorithm on the GPU. 

\begin{figure}[!h]
  \centering           
        \fbox{\includegraphics[width=0.45\textwidth]{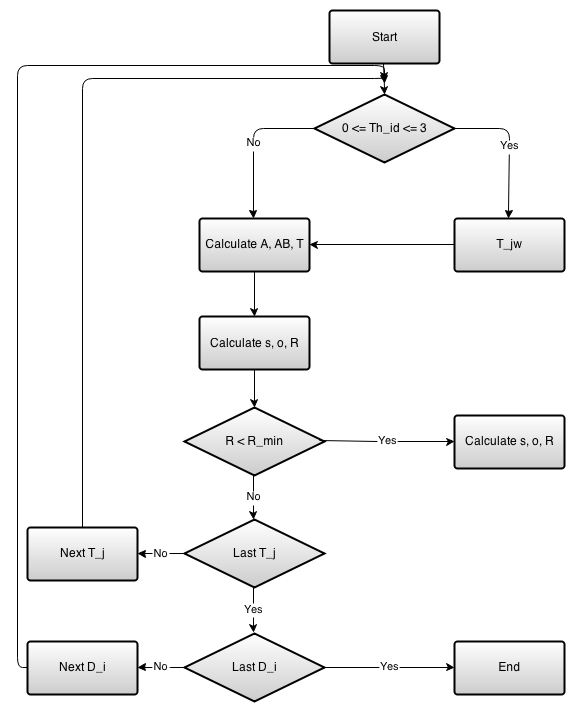}}   
        \caption{Fractal Compression Flow}         
\end{figure}


\section{Conclusion}

We have used global and texture memory to implement our CUDA program to get the difference between the performance. We have used $8\times8$, $16\times16$, and $32\times32$ blocks of kernels to test our program and found that $16\times16$ blocks were performing better.

\section{Future Directions}
In our experiment we have used black and white images from medical results. In future we wish to expand our research in other fields, e.g. traffic monitoring systems. It will be very helpful if we can process the images of vehicles from the traffic system to get the similarities among vehicles.

\appendices


\ifCLASSOPTIONcaptionsoff
  \newpage
\fi


%

\begin{IEEEbiography}
[{\includegraphics[width=1in,height=1.26in,clip,keepaspectratio]{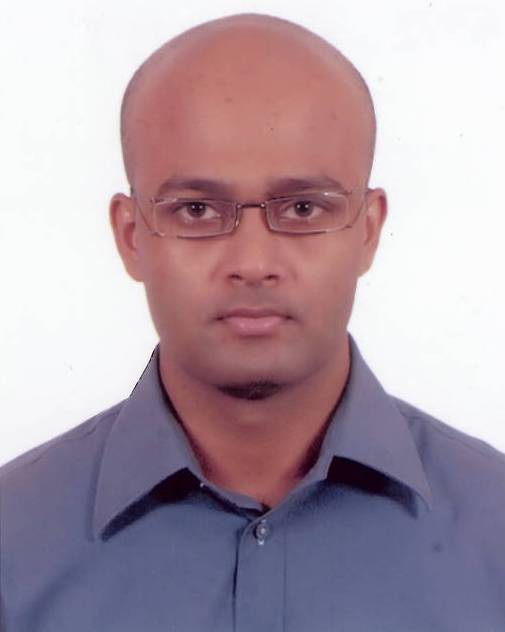}}]
{Md. Enamul Haque}
is pursuing M.Sc. in Computer Engineering at King Fahd University of Petroleum and Minerals, Saudi Arabia. He completed his B.Sc in Computer Science and IT from Islamic University of Technology (IUT), OIC. His research interest includes Autonomous sensor systems, Wireless ad-hoc networks, Computer vision and Image processing. Previously he worked as a Software Engineer in Grameenphone limited, one of the leading telecommunication company in Bangladesh.
\end{IEEEbiography}

\begin{IEEEbiography}
[{\includegraphics[width=1in,height=1.26in,clip,keepaspectratio]{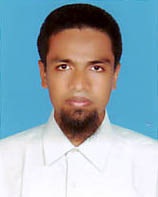}}]
{Abdullah Al Kaisan }
is currently studying M.Sc (CSE) in KFUPM. His area of interest is in computer networking. Before coming to KFUPM, he served almost four years in Rahimafrooz Bangladesh Ltd as a senior officer in SAP automation \& communication module. He completed his B.Sc in Computer Science and Engineering from Khulna University of Engineering and Technology (KUET), Bangladesh.
\end{IEEEbiography}

\begin{IEEEbiography}
[{\includegraphics[width=1in,height=1.26in,clip,keepaspectratio]{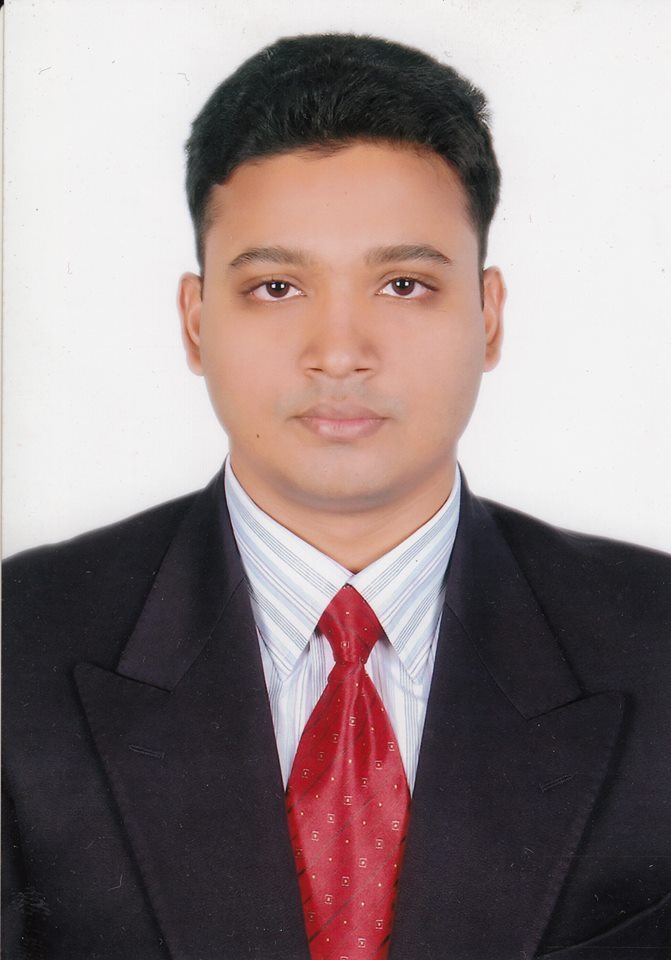}}]
{Mahmudur R Saniat}
is pursuing an Erasmus Master in Service Engineering jointly delivered by University of Stuttgart, University of Crete and Tilburg University. He completed his B.Sc in Computer Science and IT from Islamic University of Technology (IUT), OIC. His research interest include Image Processing, Process Mining, Process Re-engineering, Cloud Based Service Enabling etc. Previously he worked as an IT Executive in Jamuna Bank of Bangladesh.
\end{IEEEbiography}

\begin{IEEEbiography}
[{\includegraphics[width=1in,height=1.26in,clip,keepaspectratio]{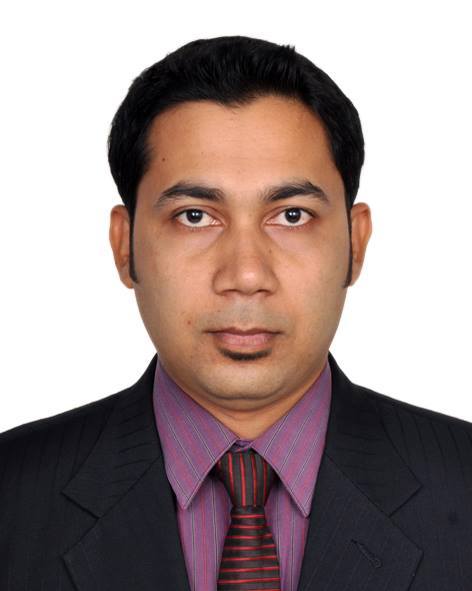}}]
{Aminur Rahman} 
is currently working as an Application Engineer in Banglalink Digital Communications Ltd, Bangladesh. He completed his B.Sc in Computer Science and IT from Islamic University of Technology (IUT), OIC. His research interest includes Computer vision and Image processing, Wireless Sensor network, AI, Natural Language Processing, Cryptography and Software Engineering. Previously he worked as programmer and Application Developer in several places in Bangladesh and Saudi Arabia. 
\end{IEEEbiography}

\vfill
\vfill







\end{document}